\documentclass[aps,preprint,nofootinbib]{revtex4}%
\usepackage{amsfonts}
\usepackage{amsmath}
\usepackage{amssymb}
\usepackage{lipsum}
\usepackage{xcolor}
\usepackage{amssymb}
\usepackage{amsmath}
\usepackage{slashed}
\usepackage{float}
\usepackage{physics}
\usepackage{graphicx}
\usepackage{hyperref}
\hypersetup{colorlinks}
\usepackage{cancel}
\usepackage[utf8]{inputenc}
\usepackage{stackrel}

\newcommand\blfootnote[1]{%
  \begingroup
  \renewcommand\thefootnote{}\footnote{#1}%
  \addtocounter{footnote}{-1}%
  \endgroup
}
\usepackage{float}
\usepackage{graphicx}%
\providecommand{\U}[1]{\protect\rule{.1in}{.1in}}

\begin{document}
 
\title{General Relativity from Einstein-Gauss-Bonnet gravity}

\author{$^{1\wedge}$Fabrizio Canfora, $^{2 , \star}$Adolfo Cisterna, \\ $^{3,*}$Sebasti\'an Fuenzalida, $^{4,\dagger}$Carla Henr\'iquez-B\'aez, $^{4,\ddagger}$Julio Oliva}

\affiliation{$^{1}$ Centro de Estudios Cient\'ificos (CECs), Casilla 1469, Valdivia, Chile}

\affiliation{$^{2}$Sede Esmeralda, Universidad de Tarapacá, Av. Luis Emilio Recabarren 2477, Iquique, Chile}

\affiliation{$^{3}$Departamento de F\'isica, Universidad T\'ecnica Federico Santa Mar\'ia, Casilla 110-V, Valpara\'iso, Chile}

\affiliation{$^{4}$Departamento de F\'isica, Universidad de Concepci\'on, Casilla 160-C, Concepci\'on, Chile}

\begin{abstract}

In this work we show that Einstein gravity in four dimensions can be consistently obtained from the compactification of a generic higher curvature Lovelock theory in dimension $D=4+p$, being $p\geq1$. The compactification is performed on a direct product space $\mathcal{M}_D=\mathcal{M}_4\times\mathcal{K}^p$, where $\mathcal{K}^p$ is a Euclidean internal manifold of constant curvature. The process is carried out in such a way that no fine tuning between the coupling constants is needed. The compactification requires to dress the internal manifold with the flux of suitable $p$-forms whose field strengths are proportional to the volume form of the internal space. We explicitly compactify Einstein-Gauss-Bonnet theory from dimension six to Einstein theory in dimension four and sketch out a similar procedure for this compactification to take place starting from dimension five. Several black string/p-branes solutions are constructed, among which, a five dimensional asymptotically flat black string composed of a Schwarzschild black hole on the brane is particularly interesting. Finally, the thermodynamic of the solutions is described and we find that the consistent compactification modifies the entropy by including a constant term, which may induce a departure from the usual behavior of the Hawking-Page phase transition. New scenarios are possible in which large black holes dominate the canonical ensamble for all temperatures above the minimal value.\blfootnote{ canfora@cecs.cl adolfo.cisterna.r@mail.pucv.cl sebastian.fuenzalidg@usm.cl carlalhenriquez@udec.cl juoliva@udec.cl}
\end{abstract}

\maketitle

\section{Introduction}

The original Kaluza-Klein scheme \cite{Kaluza,Klein} provides, via dimensional reduction to four dimensions, a consistent method to unify gravity and electromagnetism by starting from a purely geometrical higher dimensional theory \cite{Overduin:1998pn}. In higher dimensions, it is assumed that only gravity exists, described by a five dimensional Einstein-Hilbert action, that gives rise to the electromagnetic force once the spacetime is properly compactified to dimension four. 
Two fundamental assumptions are made: First, gravity is described by Einstein General Relativity (GR) and second, the higher dimensional spacetime is empty. 
Despite this conceptual appealing of the Kaluza-Klein framework, some subtleties emerge given by two experimental signatures that are in tension with the simplest approaches of compactification: the small, positive value of the four dimensional cosmological constant that we observe today and the highly accurate particle physics experiments whose consistency requires strict bounds on the size of the extra dimensions. In the cosmological context, this procedure has made it possible to find solutions of the Friedmann-Robertson-Walker class \cite{Castillo-Felisola:2016kpe}. \\
In addition, simple compactifications over direct product spaces, namely consistent compactifications with vanishing gauge and dilaton fields, where the internal manifold is of constant curvature, usually suffer from incompatibilities at the level of the field equations \cite{Duff:1986hr}. A concrete example to see this, is the compactification of Einstein theory on a $D=d+p$ dimensional spacetime of the form $\mathcal{M}_D=\mathcal{M}_d\times\mathcal{K}^p$, where $\mathcal{M}_d$ is a $d$-dimensional spacetime and $\mathcal{K}^p$ an internal $p$-dimensional Euclidean manifold of constant curvature.
Compatibility of the field equations will imply a vanishing of the Ricci tensor of the internal manifold, which if assumed of constant curvature, implies that it must be locally flat and consequently, its isometry group can only be Abelian. Isometries of the internal manifold govern the symmetry group of the compactified interaction and in consequence within this ansatz of cylindrycal compactification, GR it is only unifiable with electromagnetism.   
On the other hand, to accommodate a small non-vanishing four dimensional cosmological constant, it is then mandatory to include an internal manifold with a non-vanishing curvature which forces the radius of the extra dimensions to be large. In consequence, to properly compactify Einstein theory over internal manifolds constructed with small dimensions, with or without a small cosmological constant, it is necessary to either give up on the emptiness of the higher dimensions or to generalize the geometric structure of the theory by including well-possed higher curvature terms into the gravitational action. Including matter fields in higher dimensions seems to go against the very idea of geometrization of interactions behind Kaluza-Klein compactifications. As stated in the 80's, this would be still consistent if there is a guiding principle that fixes the matter content in higher dimensions in a relatively unique manner. Supergravity achieves this goal in eleven dimensions, since its bosonic matter content in uniquely defined by a metric, and a fundamental three-form, which cannot be further coupled to other matter multiplets \cite{Duff:1986hr}. \\
On the other hand, at a practical level, including higher dimensional matter fields has been shown to be fruitful to compactify Einstein theory. It is known that dressing the internal manifold with fundamental $(p-1)$-forms $A_{[p-1]}$, with a field strength $F_{[p]}=dA_{[p-1]}$ proportional to the volume form of such space, renders possible compactifications of Einstein gravity \cite{Freund:1980xh,RandjbarDaemi:1982hi}. In fact, the $p$-form charge allows for a non-trivial curvature of the internal manifold while at the same time accommodating a small cosmological constant.  This dressing protocol is also widely applied in the construction of black holes with $p$-form fields depending on the horizon coordinates which support a black hole horizon from collapsing when deviating from bald solutions \cite{Bardoux:2012aw} (see also \cite{Edery:2015wha}, \cite{Edery:2018jyp}) \\
In higher dimensions, higher curvature terms when combined properly in such a way that the resulting theory of gravity respect general covariance and possesses second order field equations, give rise to Lovelock theory, the natural generalization of Einstein gravity in higher dimensions \cite{Lovelock:1971yv}. This theory corresponds to an infinite series of higher curvature terms of order $k$, which are non-trivial when $k<D/2$. In this way, on top of a cosmological term, the Ricci scalar is the first term of the series and the only one that it is non-trivial in dimension four. 
The first dynamical correction of Lovelock gravity to the Einstein-Hilbert action appears in dimension five and is given by a precise combination of quadratic curvature terms known as the Gauss-Bonnet density \cite{Garraffo:2008hu}, which also appears as an $\alpha'$-correction to GR in string theory \cite{Zwiebach:1985uq}. It is natural to wonder, how this model, dubbed Einstein-Gauss-Bonnet (EGB) gravity, compactifies on direct product spacetimes. 
Einstein-Gauss-Bonnet gravity with a cosmological constant can be easily compactified over an internal manifold of non-vanishing constant curvature of dimension $p$ to a spacetime of dimension $d\geq 5$. 
The presence of the Gauss-Bonnet density renders the compactification trivial but at the price of both, tuning the cosmological constant with the Gauss-Bonnet coupling and forcing the internal manifold to have an hyperbolic structure, i.e a negative constant curvature. 
The compactified theory becomes ill-behaved, the Killing vectors of the hyperbolic internal manifold are not globally defined and then it is not possible to accommodate the unification of gravity with any other interaction. 
On might be tempted to dress the internal manifold with $p$-forms, however we have shown in \cite{Cisterna:2020kde} that this only eliminates the tuning of the cosmological constant with the Gauss-Bonnet coupling, but the internal manifold remains hyperbolic. To solve these problems it is mandatory to include higher dimensional matter in the shape of $p$-forms non-minimally coupled to the curvature tensor, keeping the second order character of the theory as a guiding principle. Such an interaction, first described by Horndeski in \cite{Horndeski:1976gi} and later generalized in \cite{Feng:2015sbw} provides the only non-minimally coupled gauge invariant electrodynamics with second order field equations that when going to flat spacetime reduces to Maxwell equations. This model renders possible a generic compactification of Einstein-Gauss-Bonnet gravity on an internal manifold of positive constant curvature \cite{Cisterna:2020kde}, a fact that is extendable to any Lovelock theory. Regarding these compactifications, a natural concern arises: Whether or not a given Lovelock theory can be compactified to Einstein gravity in dimension four? In four dimensions Lovelock densities are either topological or identically zero, nevertheless, after the compactification, traces of these geometrical entities survive at the level of the field equations, rendering the resulting theory not compatible. In \cite{Canfora:2008iu} it has been shown that Lovelock gravity can be compactified to Einstein theory in dimension four as long as the theory at least includes the cubic term of the Lovelock series, and in consequence with an internal manifold of dimension $p\geq 3$. The price to pay for this spontaneous compactification, beyond the fact that it exists starting from seven dimensions, is that the cosmological constant and the Gauss-Bonnet and cubic the Lovelock coupling are related to each other in a manner that does not lead to a symmetry enhancement, and therefore, such relation it is not expected to survive quantum corrections.\\ In this paper we show how any Lovelock theory, in particular Einstein-Gauss-Bonnet theory, can be compactified to four dimensional Einstein gravity starting from five or six dimensions and without imposing any relation among the model parameters, namely for a generic Lovelock theory. To accomplish this we employ the same strategy of \cite{Cisterna:2020kde} and we show that there is no need to consider any interaction of order higher than two in the curvature. \\
The article is organized as follows: Section II is devoted to introduce our model and the corresponding field equations. In Section III we show how to use our model to compactify Einstein-Gauss-Bonnet theory to Einstein gravity in dimension four by considering a two-dimensional internal manifold of constant curvature.
In section IV we provide a method for this compactification to take place starting from five dimensions. 
Section V is destined to study the thermodynamic features of the corresponding anti-de Sitter black brane solutions of section III and IV. We conclude in Section VI. 
 
\section{Theory and field equations}

We start by considering the Lovelock Lagrangian of order $k$ in $D$ dimensions
\begin{align}
\mathcal{L}_{Lovelock}[g]&=\sum^{[(D-1)/2]}_{k=0}\alpha_k\mathcal{L}^{k},\label{eq:general-action-geometry}
\end{align}
where the $\alpha_k$ are the Lovelock couplings and $\mathcal{L}^{k}$ the Euler densities defined by
\begin{align}
\mathcal{L}^{k}=\frac{1}{2^{k}}\delta^{A_1\cdots A_{2k}}_{B_1\cdots B_{2k}}R^{B_1B_2}{}{}_{A_{1}A_{2}}\cdots R^{B_{2k-1}B_{2k}}{}{}_{A_{2k-1}A_{2k}},
\end{align}
being $\delta^{A_1\cdots A_{2k}}_{B_1\cdots B_{2k}}$ the anti-symmetric generalized Kronecker delta. Each Euler density will contribute non-trivially to the field equations for $k\leq \left[\frac{(D-1)}{2}\right]$, otherwise they are either topological or identically zero. Notice that we have parameterized the sum in \eqref{eq:general-action-geometry} in such a way that in even dimension the topological Euler density is not present.\\
We will perform compactifications of (\ref{eq:general-action-geometry}) from a $D=d+p$ dimensional direct product spacetime $\mathcal{M}_D$ to a $d$-dimensional spacetime $\mathcal{M}_d$, being the internal manifold $\mathcal{K}^p$ a Euclidean $p$-dimensional space of constant curvature. 
As previously explained, in order to perform the compactification to Einstein gravity in dimension four it is necessary to include suitable non-minimal couplings between curvature tensors and $p$-form fields. The corresponding theory \cite{Feng:2015sbw} is constructed in complete analogy with the Lovelock Lagrangian, and it is built in terms of a polynomial invariant made of curvature tensors and the field strength of the corresponding $p$-form fields. We start by introducing the bi-linear combination 
\begin{align}
Z^{A_1\cdots A_p}{}_{B_1\cdots B_p}&:=F^{A_1\cdots A_p}F_{B_1\cdots B_p},\label{eq:Z-tensor}
\end{align}
where $F_{[p]}=dA_{[p-1]}$, so the new interaction can be defined as
\begin{align}
\mathcal{L}_{p-forms}\left[g, A_{[p-1]}\right]=\sum^{[(D-1)/p]}_{n=1}\sum^{[(D-np)/2]}_{k=0}\frac{\beta_{k}}{2^k(p!)^{n}}&\delta^{A_1\cdots A_{2k}C_1^{1}\cdots C_1^{p}\cdots C_n^{1}\cdots C_n^{p}}_{B_1\cdots B_{2k}D_1^{1}\cdots D_1^{p}\cdots D_n^{1}\cdots D_n^{p}}R^{B_1B_2}{}{}_{A_1A_2}\cdots R^{B_{2k-1}B_{2k}}{}{}_{A_{2k-1}A_{2k}}\nonumber\\
&\times Z^{D_1^{1}\cdots D_1^{p}}{}_{C_1^{1}\cdots C_1^{p}}\cdots Z^{D_n^{1}\cdots D_n^{p}}{}_{C_n^{1}\cdots C_n^{p}},\label{eq:general-action-matter}
\end{align}
being $\beta_k$ coupling constants.
The field equations are of second order which can be proved using the Bianchi identity satisfied by the $p$-forms 
\begin{align}
\nabla_{\left[A_1\right.}F_{\left.B_1\cdots B_p\right]}=\nabla^{\left[A_1\right.}F^{\left.B_1\cdots B_p\right]}=0\label{eq:bianchi-1}
\end{align}
as well as the following identity on the $Z$ tensor
\footnotesize
\begin{align}
\nabla_{\left[D_1\right.}\nabla^{\left[C_1\right.}Z^{\left.A_1\cdots A_p\right]}{}_{\left.B_1\cdots B_p\right]}&=\nabla_{\left[D_1\right.}F^{\left[A_1\cdots A_p\right.}\nabla^{\left.C_1\right]}F_{\left.B_1\cdots B_p\right]}+p(-1)^{p}F^{\left[A_1\cdots A_p\right.}R^{E}{}_{\left[B_1D_1\right.}{}^{\left.C_1\right]}F_{\left.B_2\cdots B_p\right]E}.\label{eq:bianchi-2}
\end{align}
\normalsize
Note that the order $k$ in \eqref{eq:general-action-matter} can be set to zero and only $Z$ tensors might appear in the theory. Such a model, dubbed quasitopological electromagnetism has been recently explored in the context of spherically symmetric black holes \cite{Liu:2019rib,Cisterna:2020rkc,Cano:2020qhy}. As a further guiding principle, we require the matter field equations to be linear on the $p$-form field, and therefore we restrict to Lagrangians that are quadratic in $F_{[p]}$ in \eqref{eq:general-action-matter} and therefore linear in $Z$ namely, we keep as non-vanishing only the contributions coming from the term with $n=1$ in \eqref{eq:general-action-matter}.\\
The metric field equations of (\ref{eq:general-action-geometry}) supplemented with (\ref{eq:general-action-matter}) consequently take the form 
\begin{align}
\sum^{[(D-1)/2]}_{k=0}\alpha_{k}E^{\left(k\right)}_{AB}-\sum^{[(D-p)/2]}_{k=0}\beta_{k}T^{\left(k, 1\right)}_{AB,p}&=0,\label{eq:field-equations-action-theory}
\end{align}
where $E^{\left(k\right)}_{AB}$ is the Lovelock tensor of order $k$ in the curvature, 
\begin{align}
E^{\left(k\right)}_{AB}&=-\frac{1}{2^{k+1}}g_{\left(A\right|C}\delta^{CA_{1}\cdots A_{2k}}_{\left|B\right)B_1\cdots B_{2k}}R^{B_1B_2}{}{}_{A_{1}A_{2}}\cdots R^{B_{2k-1}B_{2k}}{}{}_{A_{2k-1}A_{2k}}\label{eq:lovelock-tensor-p}
\end{align}
and $T^{\left(k, 1\right)}_{AB,p}$ is the corresponding energy-momentum tensor associated to (\ref{eq:general-action-matter}),
\begin{align}
T^{\left(k, 1\right)}_{AB,p}=&\frac{1}{2^{k+1}p!}g_{AB}\delta^{A_1\cdots A_{2k}C_1\cdots C_p}_{B_1\cdots B_{2k}D_1\cdots D_p}R^{B_1B_2}{}{}_{A_{1}A_{2}}\cdots R^{B_{2k-1}B_{2k}}{}{}_{A_{2k-1}A_{2k}}Z_{}^{D_1\cdots D_p}{}_{C_1\cdots C_p}\nonumber\\
&-\frac{k}{2^{k}p!}\delta^{A_1A_2\cdots A_{2k}C_1\cdots C_p}_{B_1\left(A\right|\cdots B_{2k}D_1\cdots D_p}R^{B_1}{}_{\left|B\right)A_1A_2}R^{B_3B_4}{}{}_{A_3A_4}\cdots R^{B_{2k-1}B_{2k}}{}{}_{A_{2k-1}A_{2k}}Z_{}^{D_1\cdots D_p}{}_{C_1\cdots C_p}\nonumber\\
&+\frac{2k}{2^{k}p!}\delta^{A_1\cdots A_{2k}C_1\cdots C_p}_{\left(A\right|\cdots B_{2k}D_1\cdots D_p}g_{A_2\left|B\right)}R^{B_3B_4}{}{}_{A_3A_4}\cdots R^{B_{2k-1}B_{2k}}{}{}_{A_{2k-1}A_{2k}}\nabla_{A_1}F_{}^{D_1\cdots D_p}\nabla^{B_2}F^{}_{C_1\cdots C_p}\nonumber\\
&+\frac{2pk}{2^{k}p!}\delta^{A_1\cdots A_{2k}C_1\cdots C_p}_{\left(A\right|\cdots B_{2k}D_1\cdots D_p}g_{A_2\left|B\right)}R^{B_3B_4}{}{}_{A_3A_4}\cdots R^{B_{2k-1}B_{2k}}{}{}_{A_{2k-1}A_{2k}}R^{D_1}{}_{E}{}^{B_2}{}_{A_1}Z_{}^{ED_2\cdots D_p}{}_{C_1\cdots C_p}\nonumber\\
&-\frac{p}{2^{k}p!}\delta^{A_1\cdots A_{2k}C_1\cdots C_p}_{B_1\cdots B_{2k}\left(A\right|\cdots D_p}R^{B_1B_2}{}{}_{A_{1}A_{2}}\cdots R^{B_{2k-1}B_{2k}}{}{}_{A_{2k-1}A_{2k}}Z_{\left|B\right)}{}^{D_2\cdots D_p}{}_{C_1\cdots C_p},\label{eq:energy-momentum-tensor-k-p}
\end{align}
while for the gauge field we have 
\begin{align}
\sum^{[(D-p)/2]}_{k=0}\frac{\beta_{k}}{2^{k-1}}\delta^{A_1\cdots A_{2k}C_1\cdots C_p}_{B_1\cdots B_{2k}D_1\cdots D_p}R^{B_1B_2}{}{}_{A_1A_2}\cdots R^{B_{2k-1}B_{2k}}{}{}_{A_{2k-1}A_{2k}}\nabla^{D_1}F_{C_1\cdots C_p} = 0.
\end{align}

In the next section we will address the compactification of Einstein-Gauss-Bonnet gravity from dimension $D=d+p$ to Einstein theory in dimension four, paying particular attention to the six dimensional case. 

\section{Compactifying Einstein-Gauss-Bonnet gravity to Einstein theory in dimension four}

Let us address now the compactification of Einstein-Gauss-Bonnet theory from dimension $d+p$ to Einstein gravity in dimension four, for generic values of the couplings. We will provide the general analysis for an arbitrary dimension $d$ and set $d=4$ when required. For the compactification to exists, our interaction (\ref{eq:general-action-matter}) should contain all terms up to quadratic order in the Riemann tensor. This imposes the following action principle 
\begin{align}
I\left[g,A_{\left[p-1\right]}\right]=&\int\sqrt{-g}d^{d+p}x\left(R-2\Lambda+\frac{\alpha_2}{4}\delta^{A_1\cdots A_4}_{B_1\cdots B_4}R^{B_1B_2}{}{}_{A_1A_2}R^{B_3B_4}{}{}_{A_3A_4}-\frac{1}{2p}Z_{}^{C_1\cdots C_p}{}_{C_1\cdots C_p}\right.\nonumber\\
&\left.+\frac{\beta_1}{2p!}\delta^{A_1A_2C_1\cdots C_p}_{B_1B_2D_1\cdots D_p}R^{B_1B_2}{}{}_{A_1A_2}Z_{}^{D_1\cdots D_p}{}_{C_1\cdots C_p}\right.\nonumber\\
&\left.+\frac{\beta_2}{4p!}\delta^{A_1\cdots A_4C_1\cdots C_p}_{B_1\cdots B_4D_1\cdots D_p}R^{B_1B_2}{}{}_{A_1A_2}R^{B_3B_4}{}{}_{A_3A_4}Z_{}^{D_1\cdots D_p}{}_{C_1\cdots C_p}\right).\label{eq:lag}
\end{align}
Notice that when $d+p=6$, this is the most general combination of the Lovelock-like form leading to a linear matter equation. Here the couplings $\alpha_2$ and $\beta_1$ have mass dimension $-2$, while $\beta_2$ has mass dimension $-4$. Varying the functional (\ref{eq:lag}) with respect to the metric field we obtain 
\begin{align}
G_{AB}+\Lambda g_{AB}+\alpha_2 H_{AB}&=T^{\left(0,1\right)}_{AB}+\beta_1 T^{\left(1,1\right)}_{AB}+\beta_2 T^{\left(2,1\right)}_{AB},\label{eq:field-eqs}
\end{align}
where $G_{AB}$ and $H_{AB}$ are respectively the Einstein and Gauss-Bonnet tensors
\begin{align}
G_{AB}&=R_{AB}-\frac{1}{2}g_{AB}R,\\
H_{AB}&=2RR_{AB}-4R_{AC}R^{C}{}_{B}-4R^{CD}R_{ACBD}+2R_{ACDE}R_{B}{}^{CDE}-\frac{1}{2}g_{AB}\mathcal{GB},
\end{align}
with $\mathcal{GB}=R^2-4R_{AB}R^{AB}+R_{ABCD}R^{ABCD}$ corresponding to the Gauss-Bonnet density.
In addition, the first energy-momentum tensor $T^{\left(0,1\right)}_{AB}$ gives the dynamics of the minimally coupled contribution
\begin{align}
T^{\left(0,1\right)}_{AB,p}=&\frac{1}{2}Z_{B}{}{}^{C_2\cdots C_p}{}_{AC_2\cdots C_p}-\frac{1}{4p}g_{AB}Z_{}^{C_1\cdots C_p}{}_{C_1\cdots C_p},\label{eq:T0}
\end{align}
while the non-minimally coupled sectors contribute as
\begin{align}
T^{\left(1,1\right)}_{AB,p}=&\frac{1}{4p!}g_{AB}\delta^{A_1A_2C_1\cdots C_p}_{B_1B_2D_1\cdots D_p}R^{B_1B_2}{}{}_{A_1A_2}Z_{}^{D_1\cdots D_p}{}_{C_1\cdots C_p}-\frac{p}{2p!}\delta^{A_1A_2C_1\cdots C_p}_{B_1B_2\left(A\right|\cdots D_p}R^{B_1B_2}{}{}_{A_1A_2}Z_{\left|B\right)}{}^{D_2\cdots D_p}{}_{C_1\cdots C_p}\nonumber\\
&-\frac{1}{2p!}\delta^{A_1A_2C_1\cdots C_p}_{B_1\left(A\right|D_1\cdots D_p}R^{B_1}{}_{\left|B\right)A_1A_2}Z_{}^{D_1\cdots D_p}{}_{C_1\cdots C_p}+\frac{1}{p!}\delta^{A_1A_2C_1\cdots C_p}_{\left(A\right|B_2D_1\cdots D_p}g_{A_2\left|B\right)}\nabla_{A_1}F_{}^{D_1\cdots D_p}\nabla^{B_2}F^{}_{C_1\cdots C_p}\nonumber\\
&+\frac{p}{p!}\delta^{A_1A_2C_1\cdots C_p}_{\left(A\right|B_2D_1\cdots D_p}g_{A_2\left|B\right)}R^{D_1}{}_{E}{}^{B_2}{}_{A_1}Z_{}^{ED_2\cdots D_p}{}_{C_1\cdots C_p}\label{eq:T1}
\end{align}
and
\begin{align}
T^{\left(2,1 \right)}_{AB,p}=&\frac{1}{8p!}g_{AB}\delta^{A_1\cdots A_4C_1\cdots C_p}_{B_1\cdots B_4D_1\cdots D_p}R^{B_1B_2}{}{}_{A_1A_2}R^{B_3B_4}{}{}_{A_3A_4}Z_{}^{D_1\cdots D_p}{}_{C_1\cdots C_p}\nonumber\\
&-\frac{p}{4p!}\delta^{A_1\cdots A_4C_1\cdots C_p}_{B_1\cdots B_4\left(A\right|\cdots D_p}R^{B_1B_2}{}{}_{A_1A_2}R^{B_3B_4}{}{}_{A_3A_4}Z_{\left|B\right)}{}^{D_2\cdots D_p}{}_{C_1\cdots C_p}\nonumber\\
&-\frac{1}{2p!}\delta^{A_1A_2\cdots A_4C_1\cdots C_p}_{B_1\left(A\right|\cdots B_4D_1\cdots D_p}R^{B_1}{}_{\left|B\right)A_1A_2}R^{B_3B_4}{}{}_{A_3A_4}Z_{}^{D_1\cdots D_p}{}_{C_1\cdots C_p}\nonumber\\
&+\frac{1}{p!}\delta^{A_1\cdots A_4C_1\cdots C_p}_{\left(A\right|\cdots B_4D_1\cdots D_p}g_{A_2\left|B\right)}R^{B_3B_4}{}{}_{A_3A_4}\nabla_{A_1}F_{}^{D_1\cdots D_p}\nabla^{B_2}F^{}_{C_1\cdots C_p}\nonumber\\
&+\frac{p}{p!}\delta^{A_1\cdots A_4C_1\cdots C_p}_{\left(A\right|\cdots B_4D_1\cdots D_p}g_{A_2\left|B\right)}R^{D_1}{}_{E}{}^{B_2}{}_{A_1}R^{B_3B_4}{}{}_{A_3A_4}Z_{}^{ED_2\cdots D_p}{}_{C_1\cdots C_p}.\label{eq:T2}
\end{align}
By virtue of the identities (\ref{eq:bianchi-1}) and (\ref{eq:bianchi-2}) we clearly see that the energy-momentum tensors (\ref{eq:T1})-(\ref{eq:T2}) are casted in a manifestly second-order fashion. 
On the other hand, variations with respect to the gauge field deliver the following second order Maxwell like equation
\begin{align}
&(p-1)!\nabla^{D_1}F_{D_1\cdots D_p}-\beta_1\delta^{A_1A_2C_1\cdots C_{p}}_{B_1B_2D_1\cdots D_p}R^{B_1B_2}{}{}_{A_1A_2}\nabla^{D_1}F_{C_1\cdots C_p}\nonumber\\
&-\frac{\beta_2}{2}\delta^{A_1\cdots A_4C_1\cdots C_p}_{B_1\cdots B_4D_1\cdots D_p}R^{B_1B_2}{}{}_{A_1A_2}R^{B_3B_4}{}{}_{A_3A_4}\nabla^{D_1}F_{C_1\cdots C_p}=0. \label{gaugeq}
\end{align}
In order to proceed with the compactification on $\mathcal{M}_D=\mathcal{M}_d\times\mathcal{K}^p$, we consider the following direct product metric
\begin{align}
ds^{2}&=g_{AB}dx^{A}dx^{B}=\tilde{g}_{\mu\nu}\left(y\right)dy^{\mu}dy^{\nu}+\hat{g}_{ij}\left(z\right)dz^{i}dz^{j}. \label{eq:ansatz-1}
\end{align}
Here $\tilde{g}_{\mu\nu}dy^{\mu}dy^{\nu}$ stands for the $d$-dimensional spacetime manifold $\mathcal{M}_d$ while $\hat{g}_{ij}\left(z\right)dz^{i}dz^{j}$ represents a $p$-dimensional Euclidean manifold $\mathcal{K}^p$ 
\begin{align}
\hat{g}_{ij}\left(z\right)dz^{i}dz^{j}&=\frac{d\vec{z}\cdot d\vec{z}}{\left(1+\frac{\gamma}{4}\sum^{p}_{j=1}z^2_j\right)^{2}}
\end{align}
of constant curvature, i.e
\begin{equation}
\hat{R}_{ijkl}=\gamma(\hat{g}_{ik}\hat{g}_{jl}-\hat{g}_{il}\hat{g}_{jk})
\end{equation}
with $\gamma$ defining the corresponding curvature radius $R_0=|\gamma|^{-1}$. From now on quantities with a tilde are intrinsically defined on the $d$-dimensional spacetime (the brane), while quantities with a hat refer to the internal manifold of dimension $p$.
In accordance with our dressing approach, the $A_{p-1}$ gauge field lives on the internal manifold and in consequence we take it to be proportional to the volume form of $\mathcal{K}^p$
\begin{align}
\hat{F}_{i_1\cdots i_p}&=\frac{q_{m}}{\left(1+\frac{\gamma}{4}\sum^{p}_{j=1}z^2_j\right)^{p}}\hat{\epsilon}_{i_1\cdots i_p}, \label{eq:ansatz-2}
\end{align} 
where $q_m$ plays the role of a generalized magnetic charge and $\hat{\epsilon}_{i_1\cdots i_p}=\delta^{z_1 \ldots z_p }_{i_{1}\ldots i_{p}}$ is the antisymmetrized Kronecker delta. This choice immediately provides a solution of the gauge field equation (\ref{gaugeq}).\\
We are left then with the Einstein equations. As the spacetime is a direct product of two manifolds, the metric field equations split into the field equations on the brane and on the internal manifold, yielding 
\begin{align}
&\left(\alpha_2+\beta_2q^{2}_{m}p!\right)\tilde{H}_{\mu\nu}+\left[1+2\alpha_2\gamma p\left(p-1\right)+\beta_1q^{2}_{m}p!\right]\tilde{G}_{\mu\nu}\nonumber\\
&+\left[-\frac{\gamma}{2}p\left(p-1\right)+\Lambda-\frac{\alpha_2}{2}\gamma^{2}p\left(p-1\right)\left(p-2\right)\left(p-3\right)+\frac{q^{2}_{m}}{4}\left(p-1\right)!\right]\tilde{g}_{\mu\nu}=0,\label{eq:brane-fe}
\end{align}
and
\begin{align}
&\left(-\frac{\alpha_2}{2}\hat{g}_{ij}+\frac{\beta_2}{2}q^{2}_{m}p!\hat{g}_{ij}\right)\tilde{\mathcal{GB}}_d+\left[-\frac{1}{2}\hat{g}_{ij}-\alpha_2\gamma\left(p-1\right)\left(p-2\right)\hat{g}_{ij}+\frac{\beta_1}{2}q^{2}_{m}p!\hat{g}_{ij}\right]\tilde{R}_d\nonumber\\
&+\left[-\frac{\gamma}{2}\left(p-1\right)\left(p-2\right)\hat{g}_{ij}+\Lambda\hat{g}_{ij}-\frac{\alpha_2}{2}\gamma^{2}\left(p-1\right)\left(p-2\right)\left(p-3\right)\left(p-4\right)\hat{g}_{ij}-\frac{q^{2}_{m}}{4}\left(p-1\right)!\hat{g}_{ij}\right]=0.\label{eq:internal-fe}
\end{align}
When compactifying, most incompatibilities emerge from these field equations. In consequence to further analyze a possible incompatibility we trace both equations, obtaining  
\begin{align}
&\left(\alpha_2+\beta_2q^{2}_{m}p!\right)\left(4-d\right)\tilde{\mathcal{GB}}_d+\left[1+2\alpha_2\gamma p\left(p-1\right)+\beta_1q^{2}_{m}p!\right]\left(2-d\right)\tilde{R}_d+\left[-\gamma dp\left(p-1\right)+2d\Lambda\right.\nonumber\\
&\left.-\alpha_2\gamma^{2}dp\left(p-1\right)\left(p-2\right)\left(p-3\right)+\frac{q^{2}_{m}}{2}d\left(p-1\right)!\right]=0,\label{eq:trace-brane}
\end{align}
and
\begin{align}
&\left(-\alpha_2 p+\beta_2q^{2}_{m}pp!\right)\tilde{\mathcal{GB}}_d+\left[-p-2\alpha_2\gamma p\left(p-1\right)\left(p-2\right)+\beta_1q^{2}_{m}pp!\right]\tilde{R}_d+\left[-\gamma p\left(p-1\right)\left(p-2\right)+2p\Lambda\right.\nonumber\\
&\left.-\alpha_2\gamma^{2}p\left(p-1\right)\left(p-2\right)\left(p-3\right)\left(p-4\right)-\frac{q^{2}_{m}}{2}p\left(p-1\right)!\right]=0,\label{eq:trace-internal}
\end{align}
To demonstrate the net effect of the inclusion of our interaction (\ref{eq:general-action-matter}) we explicitly compactify from six dimensions to Einstein gravity in dimension four, i.e $d=4$ and $p=2$. Both traces then become
\begin{align}
\left(2\beta_1q^{2}_{m}+4\gamma\alpha_2+1\right)\tilde{R}_{4}+\left(4\gamma-4\Lambda-q^{2}_{m}\right)&=0,\label{eq:egb-to-einstein-1}\\
\left(4\beta_2q^{2}_{m}-2\alpha_2\right)\tilde{\mathcal{GB}}_{4}+\left(4\beta_1 q^{2}_{m}-2\right)\tilde{R}_{4}+\left(4\Lambda-q^{2}_{m}\right)&=0,\label{eq:egb-to-einstein-2}
\end{align}
whose compatibility is ensured by the relations 
\begin{align}
q^{2}_{m}=\frac{\alpha_2}{2\beta_{2}},\quad  \gamma=-\frac{1}{4}\frac{24\Lambda \alpha_2 \beta_{1}\beta_{2}-8\Lambda \beta_{2}^2 +\alpha_2^2 \beta_{1}-3\alpha_2 \beta_{2} }{\beta_{2}\left(8\Lambda\alpha_2 \beta_{2}-\alpha_2^2 -4\alpha_2 \beta_{1}+4\beta_{2} \right)},\label{eq:charges-curvature-d4-p2}
\end{align}
that fix the magnetic charge, which is an integration constant, in terms of the couplings $\alpha_2$ and $\beta_2$ and fix as well the compactification radius $\gamma$. A few comments are in order. First, the presence of $\beta_2$, that controls a $k=2$ order term in (\ref{eq:general-action-matter}) is mandatory otherwise, although $\mathcal{GB}_{4}$ is topological in $d=4$ it presence at the level of the field equations renders impossible the compactification, indeed it constraints the system in such a way that no general relativity solutions are allowed (see also \cite{Kastor:2006vw} and \cite{Giribet:2006ec}). Second, we observe that $\alpha_2$ and $\beta_2$ are related through the magnetic charge, an integration constant that can take any value, and in consequence there is no tuning among the couplings of the theory. This is in stark contrast to what happens in \cite{Canfora:2008iu}, where for the compactification to exists a cubic Lovelock interaction must be included whose coupling $\alpha_3$ is directly fixed in terms of $\alpha_2$ and $\Lambda$. Third, this compactification can be performed from dimension six due to the fact that all terms in (\ref{eq:general-action-matter}) are not topological and in consequence contribute non-trivially in the critical dimension, opposite to the cubic Lovelock interaction which is non-topological starting from dimension seven. Finally, it is interesting to notice that in a scenario in which the higher curvature terms arise as higher curvature corrections proper of an effective field theory, $\alpha_2\sim M^{-2} \sim \beta_1$ while $\beta_2\sim M^{-4}$, where $M$ is the energy scale that defines the effective approach, namely physical quantities must be expanded in the limit $M$ large. From the second relation in \eqref{eq:charges-curvature-d4-p2}, one can show that the compactification radius $R_0^2\sim M^{-2}$, and therefore the perturbative scheme is indeed consistent with having a perturbatively small, compact, extra dimension.\\
Therefore, after fixing the radius of compactification as well as the value of the magnetic charge as in \eqref{eq:charges-curvature-d4-p2}, one consistently obtains an Einstein equation induced on the brane, which can be read from \eqref{eq:brane-fe} leading to
\begin{equation}
\frac{1}{16\pi G_{\text{eff}}}\left(G_{\mu\nu}+\Lambda_{\text{eff}}g_{\mu\nu}\right)=0\ ,
\end{equation}
where we have defined the effective Newton and cosmological constants respectively as
\begin{align}
G_{\text{eff}}&=\frac{1}{32\pi}\frac{\beta_2(8\Lambda\alpha_2\beta_2-\alpha_2^2-4\alpha_2\beta_1+4\beta_2)}{(\beta_2-\alpha_2\beta_1)(\alpha_2^2+2\alpha_2\beta_1+8\Lambda\alpha_2\beta_2+2\beta_2)}\ ,\ \Lambda_{\text{eff}}=\frac{(8\Lambda\beta_2-\alpha_2)}{16(\beta_2-\alpha_2\beta_1)} \ , \label{GeffyLambdaeff}
\end{align}
The field equations are now solvable, for example, by the following $6$-dimensional black 2-brane (static spherically symmetric case)
\begin{equation}
ds^2=-\left(\frac{r^2}{l^2_{\text{eff}}}-\frac{\mu}{r}+K\right)dt^2+\frac{dr^2}{\left(\frac{r^2}{l^2_{\text{eff}}}-\frac{\mu}{r}+K\right)}+r^2d\Sigma^2_K+\frac{dz_1^2+dz_2^2}{[1+\frac{\gamma}{4}(z_1^2+z_2^2)]^2}\label{6D2brane}
\end{equation}
with $K$ the curvature of the transverse manifold $d\Sigma_K^2$, $\mu$ an integration constant related with ADM mass and $l_{\text{eff}}^2=-3/\Lambda_{\text{eff}}$ an effective (A)dS radius 
\begin{equation}
l_{\text{eff}}^{-2}=\frac{1}{48}\left(\frac{8 \Lambda \beta_{2} -\alpha_2}{\alpha_2\beta_{1}-\beta_{2}}\right).\label{leff42}
\end{equation}
The four dimensional brane is given then by a Schwarzschild (A)dS black hole with an effective cosmological constant, i.e an Einstein solution. From the latter equation, one can see that for perturbative, higher curvature couplings, the effective four-dimensional cosmological constant turns out to be large, which in the case of de Sitter solutions is in tension with a direct application of this scenario to late time cosmology. 

\section{The scalar case: One dimensional internal manifold}

In order to compactify Einstein-Gauss-Bonnet gravity starting from dimension five it is necessary to consider a one dimensional internal manifold, and in consequence the dressing must be given by $0$-forms, i.e scalar fields. This kind of dressing, has been successfully applied in order to construct homogenous anti-de Sitter black string/p-branes in general relativity and Lovelock theories \cite{Cisterna:2017qrb,Cisterna:2018jsx,Cisterna:2018mww,Arratia:2020hoy,Cisterna:2019scr}. Analogously to the case we have previously faced, in which the fundamental forms were proportional to the volume form of the internal manifold, the scalars are required to be linear in the internal manifold coordinate, providing in this way an immediate solution to the corresponding Klein-Gordon equation. Our interaction (\ref{eq:general-action-matter}) is modified in such a manner that now the curvature tensors are coupled with a specific combination of first derivatives of the scalars. We then consider
\begin{equation}
\mathcal{L}_{scalar}[g,\phi]=-\frac{1}{2^{2k+1}}\delta^{CA1...A_{2k}}_{DB1...B_{2k}}R^{B1B2}{}{}_{A1A2}...R^{B_{2k-1}B_{2k}}{}{}_{A_{2k-1}A_{2k}}\nabla_C\phi^{i}\nabla^D\phi^{i}
\end{equation}
which is equivalent to a generalized non-minimal kinetic coupling controlled by the Lovelock tensor of order $k$
\begin{equation}
\mathcal{L}_{scalar}[g,\phi^{i}]=-\frac{1}{2^{2k+1}}E^{CD}_{(k)}\nabla_C\phi^{i}\nabla_D\phi^{i}
\end{equation}
In complete analogy with the previous section we truncate our theory up to order $k=2$, then our action principle is written as 
\begin{align}
I\left[g,\phi,\psi \right]=&\int\sqrt{-g}d^{5}x\left[R-2\Lambda+\frac{\alpha_2}{4}\delta^{A_1\cdots A_4}_{B_1\cdots B_4}R^{B_1B_2}{}{}_{A_1A_2}R^{B_3B_4}{}{}_{A_3A_4}-\frac{1}{2}g^{AB}\nabla_A\phi\nabla_B\phi\right.\nonumber\\
&\left.+\frac{\beta_1}{8}G^{AB}\nabla_A\phi\nabla_B\phi+\frac{\gamma_1}{64}H^{AB}\nabla_A\phi\nabla_B\phi -\frac{1}{2}g^{AB}\nabla_A\psi\nabla_B\psi +\frac{\beta_{2}}{8}G^{AB}\nabla_A\psi\nabla_B\psi \right. \nonumber \\ 
&\left. +\frac{\gamma_2}{64}H^{AB}\nabla_A\psi\nabla_B\psi \right]\label{eq:lagscalar}
\end{align}
where, in order to avoid any fine tuning between the couplings, one has to introduce at least two different scalar fields (as the computation here below shows). This is not necessary in the previous six dimensional case because the curvature of the internal manifold provides an extra scale that prevents the appearance of such relations. \\
The resulting theory belongs to a sector of Horndeski gravity, the most general scalar-tensor theory with second order field equations \cite{Horndeski:1974wa}. The kinetic coupling to the Einstein tensor corresponds to a quadratic sector of Horndeski gravity in dimension four while the coupling with the Gauss-Bonnet tensor appears naturally in dimension greater or equal than five. Taking the variation with respect to the metric, we obtain
\begin{align}
G_{AB}+\Lambda g_{AB}+\alpha_2 H_{AB}&=T^{\left(0\right)}_{\phi_{i} AB}+\beta_{i} T^{\left(1\right)}_{\phi_{i} AB}+\gamma_{i} T^{\left(2\right)}_{\phi_{i} AB},\label{eq:field-eqs-sf}
\end{align}
where 
\footnotesize
\begin{align}
T^{(0)}_{\phi_{i} AB}&=\frac{1}{2}\left(\partial_{A}\phi_{i} \partial_{B}\phi_{i}-\frac{1}{2}g_{AB}\partial_{C}\phi_{i} \partial^{C}\phi_{i} \right) ,\label{eq:T0-sf} 
\end{align}
\begin{align}
T^{\left(1\right)}_{\phi_{i} AB}&=\frac{1}{2}\left(\frac{1}{2}\partial_{A}\phi_{i} \partial_{B}\phi_{i} R-2\partial_{C}\partial_{(A}\phi_{i} R_{B)}^{\, \, C}-\partial_{C}\phi_{i} \partial_{D}\phi_{i} R_{A \, \, \, B}^{\, \, \, C \, \, \, D}-\nabla_{A}\nabla^{C}\phi_{i} \nabla_{B}\nabla_{C}\phi_{i} +\nabla_{A}\nabla_{B}\phi_{i} \square \phi_{i} \right.\nonumber\\
&\left. +\frac{1}{2}G_{AB}\left(\partial \phi_{i} \right)^2-g_{AB}\left[-\frac{1}{2}\nabla^{C}\nabla^{D}\phi_{i} \nabla_{C}\nabla_{D}\phi_{i} +\frac{1}{2}\left(\square \phi_{i} \right)^2 -\partial_{C}\phi_{i} \partial_{D}\phi_{i} R^{CD} \right] \right)  \ , \label{eq:T1-sf} 
\end{align}
\begin{align}
T^{(2)}_{\phi_{i} AB}=&-\frac{1}{64}g_{AB}\nabla_C\phi_{i}\nabla^{D}\phi_{i}\delta^{CA_1\cdots A_4}_{DB_1\cdots B_4}R^{B_1B_2}{}{}_{A_1A_2}R^{B_3B_4}{}{}_{A_3A_4}+\frac{1}{32}\nabla_{C}\phi_{i}\nabla_{\left(A\right|}\phi_{i}\delta^{CA_1\cdots A_4}_{\left|B\right)B_1\cdots B_4}R^{B_1B_2}{}{}_{A_1A_2}R^{B_3B_4}{}{}_{A_3A_4}\nonumber\\
&+\frac{1}{16}\nabla_{C}\phi_{i}\nabla^{D}\phi_{i}\delta^{CA_1\cdots A_4}_{DB_1\cdots\left(A\right|}R^{B_1B_2}{}{}_{A_1A_2}R^{B_3}{}_{\left|B\right)A_3A_4}+\frac{1}{8}\nabla^{B_3}\nabla_{C}\phi_{i}\nabla_{A_3}\nabla^{D}\phi_{i}\delta^{CA_1\cdots A_4}_{DB_1\cdots \left(B\right|}R^{B_1B_2}{}{}_{A_1A_2}g_{\left|A\right)A_4}\nonumber\\
&+\frac{1}{16}\nabla_{C}\phi_{i}\nabla^{E}\phi_{i}\delta^{CA_1\cdots A_4}_{DB_1\cdots \left(B\right|}R^{B_1B_2}{}{}_{A_1A_2}R_{A_3E}{}{}^{B_3D}g_{\left|A\right)A_4}\ . \label{eq:T2-sf}
\end{align}
\normalsize
where $\phi_{i}$ stands for the scalar fields and $i=1,2$.   
On the other hand, the scalar fields fulfil
\begin{align}
\left( g^{AB} -\frac{\beta_{i}}{4}
G^{AB}-\frac{\gamma_{i}}{32} H^{AB}\right)\nabla_A\nabla_{B}\phi^{i}&=0.
\end{align}
To start with, we take a five dimensional product space $\mathcal{M}_D=\mathcal{M}_4\times\mathbb{R}$ whose metric along the lines of (\ref{eq:ansatz-1}) can be written as
\begin{align}
ds^2=d\tilde{s}_{4}^2+dz^2 \label{eq:ansatz-sf} \ .
\end{align}
As we said, each scalar field is linear in the internal manifold coordinate 
\begin{equation}
\phi(z)=\lambda_0 z, \hspace{0.5cm}\psi(z)=\lambda_1 z, 
\end{equation}
being $\lambda_0$ and $\lambda_1$ integration constants, the scalar charges, and they immediately solve each corresponding Klein-Gordon equation. Computing the field equations on the brane and on the extra coordinate and tracing them, we get
\begin{align}
\left(4\Lambda + \lambda_0^2+\lambda_{1}^2 \right)-\left(1-\frac{\beta_1}{4}\lambda_{0}^2-\frac{\beta_2}{4}\lambda_{1}^2\right)\tilde{R}&=0
\end{align}
and
\begin{align}
\left(-2\Lambda+\frac{ \lambda_0^2}{2}+\frac{ \lambda_1^2}{2}\right)+\left(1+\frac{\beta_{1}}{4}\lambda^{2}_{0}+\frac{\beta_{2}}{4}\lambda^{2}_{1}\right)\tilde{R}+\left(\alpha_2-\frac{\gamma_{1}}{8}\lambda^{2}_{0}-\frac{\gamma_{2}}{8}\lambda^{2}_{1}\right)\tilde{\mathcal{GB}}=0.
\end{align}
Compatibility will then require 
\begin{equation}
\lambda_0^2=\frac{8\alpha_2 -\gamma_{2}\lambda_{1}^2}{\gamma_1}, \hspace{0.5cm} \Lambda=-\frac{1}{4}\frac{\left(\gamma_{1}\lambda_{1}^2-\gamma_{2}\lambda_{1}^2+8\alpha_{2} \right)\left(-\beta_{1}\gamma_{2}\lambda_{1}^2+\beta_{2}\gamma_{1}\lambda_{1}^2+8\alpha_{2}\beta_{1} +12 \gamma_{1}\right)}{\gamma_{1}\left(-3\beta_{1}\gamma_{2}\lambda_{1}^2+3\beta_{2}\gamma_{1}\lambda_{1}^2+24 \alpha_{2}\beta_{1}+4\gamma_{1} \right)}. \label{comp}
\end{equation}
For an spherically symmetric ansatz on the brane, solving Einstein equations delivers
\begin{equation}
ds^2=-\left(\frac{r^2}{l^2_{\text{eff}}}-\frac{\mu}{r}+K\right)dt^2+\frac{dr^2}{\left(\frac{r^2}{l^2_{\text{eff}}}-\frac{\mu}{r}+K\right)}+r^2d\Sigma_K^2+dz^2 \label{5DBString}
\end{equation}
where the effective cosmological constant, by means of (\ref{comp}), is given by
\begin{equation}
l^{-2}_{\text{eff}}=\frac{1}{6}\frac{\left(\gamma_{1}-\gamma_{2}\right)\lambda_{1}^2+4\Lambda \gamma_{1}+8\alpha_{2}}{\left(-\beta_{1}\gamma_{2}+\beta_{2}\gamma_{1}\right)\lambda_{1}^2+8\alpha_{2}\beta_{1}-2\gamma_{1}}\ .\label{leff48}
\end{equation}
It is interesting to note that, from here, it is possible to obtain a Schwarzschild black string, i.e a five dimensional black string solution to (\ref{eq:field-eqs-sf}) in which the four dimensional brane is given by a Schwarzschild black hole. This might be done by considering the case in which $\Lambda=\lambda_0=0$. For such a case, even the $\beta_1$ interaction maybe neglected and the result is nothing else than the standard black string
\begin{equation}
ds^2=-\left(K-\frac{\mu}{r}\right)dt^2+\frac{dr^2}{\left(K-\frac{\mu}{r}\right)}+r^2d\Sigma_K^2+dz^2 \label{BSS}
\end{equation}
Here we observe the evident relation between black strings and compactifications on direct product spaces. The original five dimensional Schwarzschild black string exists because general relativity admits cylindrical compactifications of the form on $\mathcal{M}_D=\mathcal{M}_4\times\mathbb{R}(\mathcal{S}^1$). This solution is particularly interesting because in five dimensions these configurations suffers from a long wavelength instability, the Gregory-Laflamme instability \cite{Gregory:1993vy}, revealing the weakness of the cosmic censorship in higher dimensions \cite{Lehner:2010pn}. One might conjecture that this is related with the fact that the Schwarzschild black hole is a solution of general relativity, and that when uplifting the solution to a five dimensional black string we are still under the domain of the same theory, namely, solving Einstein field equations in dimension five. However, in five dimensions Einstein theory is not the most general gravitational theory we have at hand, in fact, Einstein-Gauss-Bonnet gravity plays that role. By means of our compactification procedure, we have been able to construct a black string in five dimensional Einstein-Gauss-Bonnet gravity (\ref{5DBString},\ref{BSS}) which precisely represent an Einstein solution on the four dimensional brane. In consequence it would be appealing to study its mechanical stability at least in the linear regime.

\section{Thermodynamic quantities}

We address now the thermodynamic analysis of the $6$-dimensional black $2$-brane solution (\ref{6D2brane})\footnote{The $5$-dimensional black string (\ref{5DBString}) follows the same procedure.}. The black $2$-brane temperature is obtained, as usual, by requiring a smooth Euclidean continuation of the solution, then 
\begin{align}
T &=\frac{r_+}{2\pi l_{\text{eff}}^2}+\frac{\mu(r_+)}{4\pi r_+^2}\ ,\label{eq:temperature-d6-brane}
\end{align}
where $r=r_{+}$ denotes the location of the black hole horizon and the effective AdS length is given in \eqref{leff42}. 
On the other hand, by means of Wald's formula \cite{Wald:1993nt} we obtain that the entropy density is
\begin{equation}
s=\frac{S}{\mathrm{Vol}\left[\mathcal{K}^2\right]}=\frac{\sigma r_+^2}{4G_{\text{eff}}}+32\pi\sigma K \alpha_2\ , \label{eq:entropy-d6-brane}
\end{equation}
with $G_{\text{eff}}$ defined in \eqref{GeffyLambdaeff}. Here $\mathrm{Vol}[\mathcal{K}^2]$ represents the volume of the $2$-dimensional internal manifold $\mathcal{K}^2$, while $\sigma$ corresponds to the volume of the transverse manifold of the four dimensional black hole on the brane. It is interesting to notice that there are two types of contributions to the black hole entropy. Since the effective theory that dictates the dynamics of the four dimensional manifold is GR, the entropy of the black hole has a contribution that goes as $r_+^2$, namely as the area of the event horizon. This contribution acquires corrections from all the higher curvature matter couplings, and can be written as $A/(4G_{\text{eff}})$. It is also worth pointing out that there is an extra, universal constant contribution to the entropy, which comes from the presence of the higher dimensional Gauss-Bonnet term (see also \cite{Cai:2001dz}).\\
Using the first law of black hole thermodynamic, $dm=Tds$, we obtain that the constant $\mu$ is identified to the mass of the solution by

\begin{equation}
m=\frac{\sigma r_+(r_+^2+K l_{\text{eff}}^2)}{8G_{\text{eff}}l_{\text{eff}}^2\pi}=\frac{\mu\sigma}{8\pi G_{\text{eff}}}\ , \label{eq:mass-d6-brane}
\end{equation}

where $r_+=r_+(\mu)$. We observe that there is no gap, i.e the mass density goes to zero in the limit $r_{+}\rightarrow 0$. As expected, the Kretschmann invariant for (\ref{6D2brane}) as $r\rightarrow 0$ behaves as
\begin{align}
R_{ABCD}R^{ABCD}=\mathcal{O}(r^{-6})\label{eq:invariant-6-brane}, 
\end{align}
revealing the existence of a curvature singularity located at the origin.
It is very interesting to see the effect of the constant term in the entropy, induced by the higher curvature terms in higher dimensions, which lead to a consistent dimensional reduction thanks to the non-minimally coupled Maxwell field. Considering the temperature, entropy and mass, respectively given in \eqref{eq:temperature-d6-brane}, \eqref{eq:entropy-d6-brane} and \eqref{eq:mass-d6-brane}, one finds the following expression for the Helmholtz free energy
\begin{equation}
\frac{F\left(r_+\right)}{\mathrm{Vol}\left[\mathcal{K}^2\right]}=\frac{r_+}{4G_{\text{eff}}}\left(1-\frac{r_+^2}{l_{\text{eff}}^2}\right)-32\pi\alpha_2\left(\frac{3r_+}{l_{\text{eff}}^2}+\frac{1}{r_{+}}\right)\ ,
\end{equation}
where we have focused on the spherically symmetric case, namely, we have fixed $\sigma=4\pi$ and $K=1$.
Since the temperature is not modified with respect to that of Schwarzschild-AdS, it takes a minimum value, at which the free energy reduces to
\begin{equation}
\frac{F\left(T_{\text{min}}\right)}{\mathrm{Vol}\left[\mathcal{K}^2\right]}=2\sqrt{3}l_{\text{eff}}\left(\frac{1}{36 G_{\text{eff}}}-\frac{32\pi\alpha_2}{l_{\text{eff}}^2}\right)\ .
\end{equation}
Notice that $F(T_{\text{min}})$ can have either signs depending on the precise values of the Gauss-Bonnet coupling, and the presence of the effective cosmological and Newton constant, allows to modify the sign even when the higher curvature terms are of a perturbative nature. It is also interesting to notice that for arbitrarily small black holes the free energy goes to minus infinity due to the presence of the Gauss-Bonnet coupling $\alpha_2$, in stark contrast to the behavior of small black holes in GR whose free energy asymptotically vanish. Nevertheless the presence of the additive term in the entropy, sets a lower bound on the mass of the black holes and consequently a lower bound on their radii. This standard bound comes from considering an idealized, quasi-static process of black hole fusion where the initial black holes have masses $M_1$ and $M_2$ and, disregarding the gravitational radiation, the variation of the entropy reads
\begin{equation}
\Delta S=8\pi G_{\text{eff}}M_1M_2-128\pi^2\alpha_2-\frac{64\pi G_{\text{eff}}^3}{l_{\text{eff}}^2}M_1M_2(2M_1^2+3M_1M_2+2M_2^2)+\mathcal{O}(l_{\text{eff}}^{-4})\ ,
\end{equation}
where $\Delta S=S_{\text{final}}-S_{\text{initial}}$. Since the expression for finite $l_{\text{eff}}$ is not very illuminating, we have taken the expansion of for large $l_{\text{eff}}$. In the asymptotically flat case, the positivity of $\Delta S$ clearly imposes a lower bound on the black hole masses.\\
Our discussion is based on the original interpretation of these type of terms given in \cite{Jacobson:1993xs}. According to this interpretation, the extra term in the entropy for two far apart black holes must be added, leading to a potential violation of the law of increasing entropy in a proccess of quasi-static black hole coalesence for $\alpha_2>0$, unless the mass is bounded from below. If $\alpha_{2}<0$ no such bound appears using this argument. It is important to remark that when the GB term emerges as a higher curvature correction in string theory, $\alpha_{2}$ is positive. Nevertheless, there are two extra arguments supporting the Jacobson-Myers interpretation. Firstly, this constant term (despite its temperature independence) is actually non-vanishing only in the presence of the black hole horizon, since otherwise the domain of integration is void. Secondly, there are many physical systems which possess non-zero extensive entropy at zero temperature, as for instance the antiferromagnetic Potts model with not too large q (q=2 for the Ising model).

\begin{figure}[h]
\includegraphics[scale=0.42]{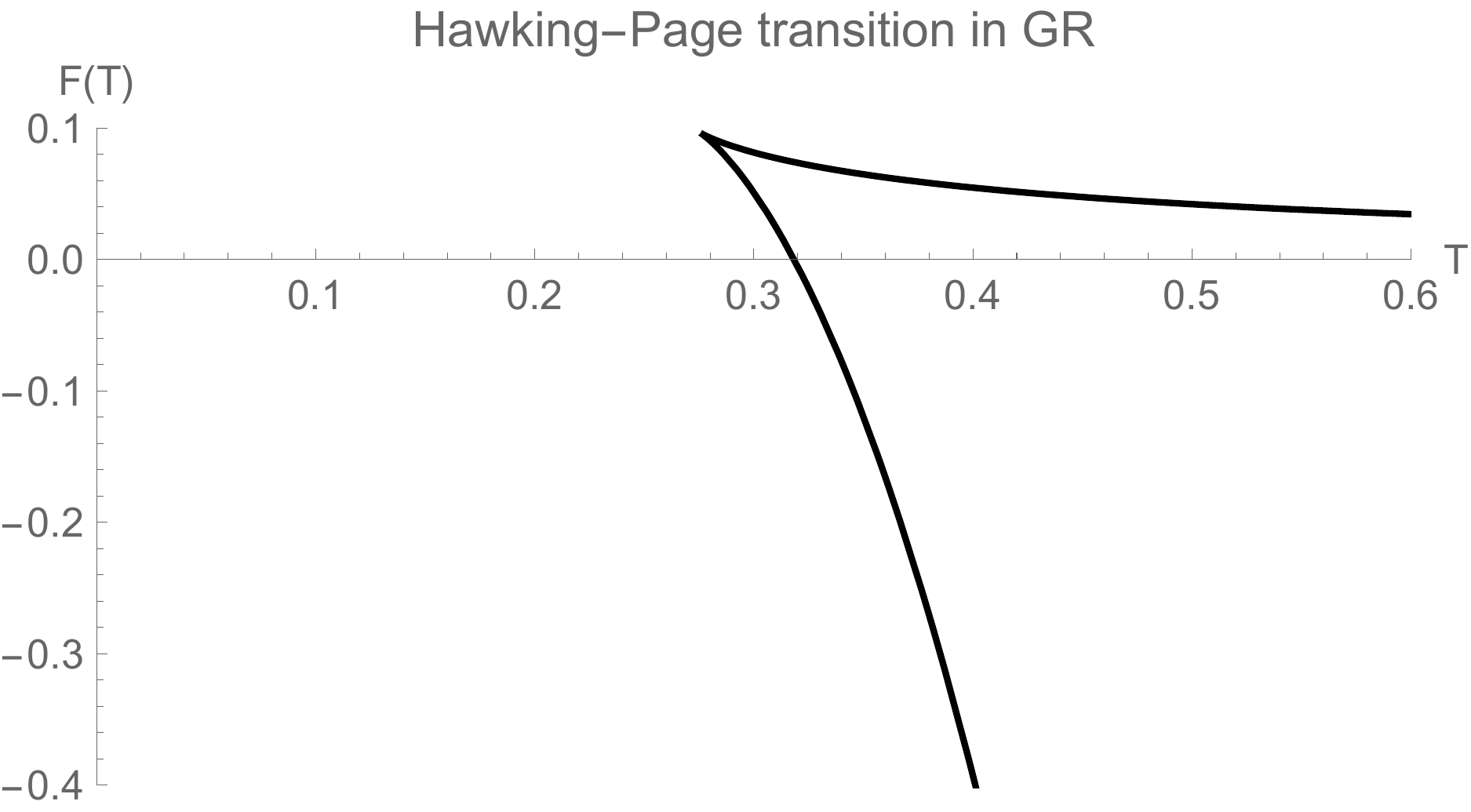}\quad\includegraphics[scale=0.4]{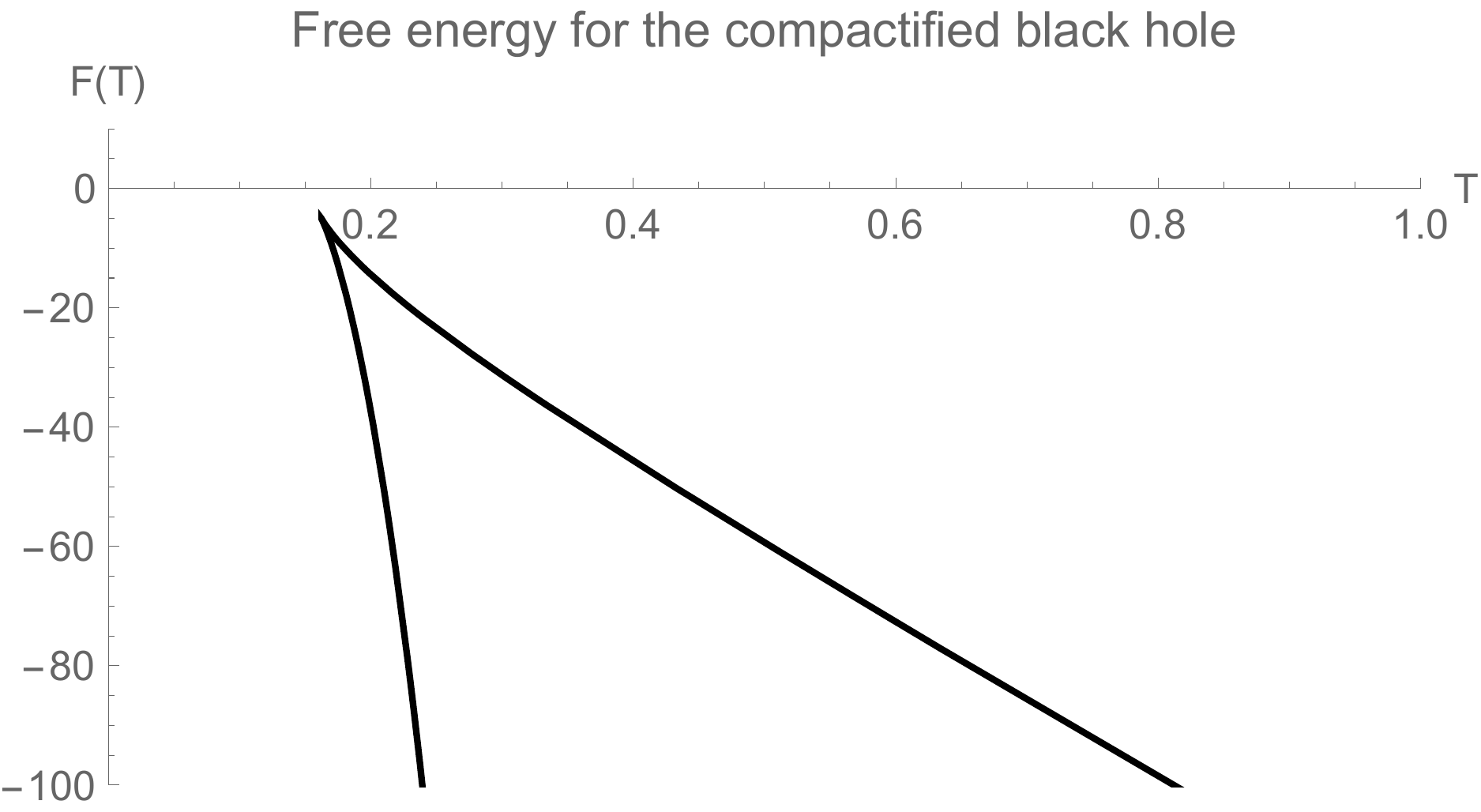}
\caption{Free energy versus temperature for Schwarzschild-AdS in General Relativity in four dimensions (left panel) and for the compactified black hole solution Schwarzschild-AdS$_4\times S^2$ (right panel). The compactified solution corresponds to the values $\alpha_2=0.1,\ \beta_1=0.2,\ \beta_2=0.05$ and $\Lambda=-1$.}
\label{fvst}
\end{figure}

Figure \ref{fvst} depicts the free energy curves for different values of the parameters. Since no other integration constant, beyond the mass is present, the appropriate ensemble is the canonical ensemble and therefore the statistically most favoured configuration is the one that minimizes the Helmholtz free energy for a given fixed temperature. It is also important to notice that the vacuum configuration, namely the solution with vanishing $F(T)$ is equal to thermal AdS$_4\times S^2$ with non-vanishing flux of the Maxwell field on the two-dimensional, compactified internal sphere. Remarkably, when $F(T_{\text{min}})$ is negative, there is no Hawking-Page transition \cite{Hawking:1982dh} and the large black holes always dominate the ensemble.\\

\section{Further comments}
We have been able to perform a dimensional reduction of EGB theory to dimension four, giving raise to General Relativity, for arbitrary values of the couplings. The consistency of the compactification relies on the presence of non-minimally coupled, magnetically charged, $p$-forms. As a guiding principle for the introduction of this matter field, we have restricted to the family of couplings introduced in \cite{Feng:2015sbw}, which mimic the structure of Lovelock theories, leading to second order field equations. The Gauss-Bonnet term, as well as other higher curvature couplings with matter fields naturally emerge as perturbative correction of fundamental theories in effective field theory scenarios (see e.g. \cite{Burgess:2007pt}). Whether or not the precise couplings considered here emerge from a low energy limit of a fundamental theory after considering the freedom of field redefinitions, is beyond our present objectives, nevertheless we have shown that in such potential scenario, the radius of compactification of the two-dimensional compact manifold considered in Section III, is indeed proportional to the inverse of the energy scale $M$, and therefore can be consistently considered as small.\\
The structure of the theories considered suggest that our results can be extended to Lovelock theories beyond the EGB Lagrangian. One might even been able to consider a given, arbitrary Lovelock theory in dimension $D$ and obtain a consistent arbitrary Lovelock theory in dimension $d=D-p$ by considering magnetically charged, non-minimally coupled $p$-forms in the family \eqref{eq:general-action-matter}. As mentioned above, in vacuum, such compactifications usually require to introduce relations between the couplings, which are not compatible with the interpretation of the higher curvature terms as perturbative corrections (for solutions of Lovelock theories in $N+1$-dimensions see \cite{Kastor:2017knv}). As one increases the dimension of the compact manifold, releasing its geometry, it is natural to expect the presence of some higher curvature constraints on its curvature. For the case of topological black holes, it was shown in \cite{Dotti:2005rc} that a new constant characterising the square of the Weyl tensor of the horizon appears on the lapse function and the effects of such constant in the thermodynamics has recently been explored in \cite{Hull:2021bry}. A simple family of manifolds that would allow to go beyond constant curvature internal spaces is the products of spheres. These manifolds will allow to introduce further parameters in the compactifications which may help when contrasting the four-dimensional obtained theory with experimental evidence. \\
Finally, as explained in Section V, the consistent compactification we have found, may lead to black holes in AdS$_4\times S^2$ which are always thermodynamically favoured with respect to the solitonic thermal background. In addition, since the Hawking-Page phase transition plays a very important role in holography \cite{Witten:1998zw}, it would be appealing to explore some holographic properties of the present compactification scenario.

\section{Acknowlegdments}
This work is partially funded by FONDECYT grants 1200022, 1210500 and 1181047. J.O. also thanks the support of Proyecto de Cooperación Internacional 2019/13231-7 FAPESP/ANID. C.H  acknowledges support from the National Agency for Research and Development (ANID) /Scholarship Program/BECA DE DOCTORADO NACIONAL/2017 - 21171394. S.F. thanks the support of Beca Doctorado USM. The Centro de Estudios Científicos (CECs) is funded by the Chilean Government through the Centers of Excellence Base Financing Program of ANID.

\end{document}